\documentclass[aps,prb,twocolumn,superscriptaddress,floatfix,longbibliography]{revtex4-2}

\usepackage{graphicx} 
\usepackage{bm} 
\usepackage{comment} 
\usepackage{endnotes} 
\usepackage{hyperref}
\hypersetup{colorlinks,linkcolor=blue,urlcolor=blue,citecolor=blue}

\usepackage{xcolor}
\definecolor{lightgray}{gray}{0.6}
\newif\ifptitle
\newif\ifpnumber
\newcounter{para}
\newcommand\ptitle[1]{\par\refstepcounter{para}
{\ifpnumber{\noindent\textcolor{lightgray}{\textbf{\thepara}}\indent}\fi}
{\ifptitle{\textbf{[{#1}]}}\fi}}

\newcommand{\Vs}{$V_\mathrm{s}$\ }
\newcommand{\Is}{$I_\mathrm{s}$\ }

\newcommand{\Vrms}{$V_\mathrm{rms}$\ }

\newcommand{\didv}{\textit{dI/dV}}
\newcommand{\ybb}{YbB\textsubscript{6}}
\newcommand{\smb}{SmB\textsubscript{6}}
\newcommand{\Ef}{\textit{E}\textsubscript{F}}

\newcommand{\Hphys}{Department of Physics, Harvard University, Cambridge, MA, 02138, USA}
\newcommand{\Heng}{School of Engineering and Applied Science, Harvard University, Cambridge, MA, 02138, USA}
\newcommand{\UCI}{Department of Physics and Astronomy, University of California, Irvine, CA 92697, USA}

\begin{document}

\title{Nanoscale Conducting and Insulating Domains on \texorpdfstring{YbB$_6$}{YbB6}}

\author{Aaron Coe}
\author{Zhi-Huai Zhu}
\author{Yang He}
\affiliation{\Hphys}
\author{Dae-Jeong Kim}
\affiliation{\UCI} 
\author{Zachary Fisk}
\affiliation{\UCI}
\author{Jason D. Hoffman}
\affiliation{\Hphys}
\author{Jennifer E. Hoffman}
\email{jhoffman@physics.harvard.edu}
\affiliation{\Hphys}
\affiliation{\Heng}

\date{\today}

\begin{abstract}
\ybb\ is a predicted topological insulator, with experimental evidence for conducting surface states of yet-unproven origin. However, its lack of a natural cleavage plane, and resultant surface-dependent polarity, has obscured its study. 
We use scanning tunneling microscopy to image the cleaved surface of \ybb, exhibiting several coexisting terminations with distinct atomic structures. 
Our spectroscopic measurements show band-bending between the terminations, resulting in both conducting and fully-gapped regions. 
In the conductive regions, we observe spectral peaks that are suggestive of van Hove singularities arising from Rashba spin-split quantum well states. 
The insulating regions rule out the possibility that \ybb\ is a strong topological insulator, while the spin-polarized conducting regions suggest possible utility for spintronic devices.
\end{abstract}

\maketitle 

\ptitle{Intro to hexaborides and TKIs} 
Rare-earth hexaborides host a diverse set of exotic properties, ranging from superconductivity to magnetism to non-trivial topology. 
Interest in hexaborides surged when \smb\ was predicted to be the first of a new type of correlated electron system, termed a topological Kondo insulator (TKI), in which protected surface states span a Kondo gap pinned around the Fermi level (\Ef) \cite{dzero_topological_2010, dzero_theory_2012}.  
Extensive experiments, including transport \cite{wolgast_low-temperature_2013,zhang_hybridization_2013, li_two-dimensional_2014, kim_surface_2013, kim_topological_2014, nakajima_one-dimensional_2016, phelan_correlation_2014, syers_tuning_2015}, angle-resolved photoemission spectroscopy (ARPES) \cite{xu_surface_2013, jiang_observation_2013, neupane_surface_2013, frantzeskakis_kondo_2013, zhu_polarity-driven_2013, denlinger_temperature_2014, min_importance_2014, xu_direct_2014} and scanning tunneling microscopy/spectroscopy (STM/STS) \cite{rosler_hybridization_2014, ruan_emergence_2014}, demonstrated the existence of two-dimensional (2D) metallic surface states on \smb\ at low temperatures. However, the topological nature of these states was unclear due to the multiple atomic terminations with complicated polar structure \cite{zhu_polarity-driven_2013, rosler_hybridization_2014, ruan_emergence_2014, matt_consistency_2020}. 
Recently, quasiparticle interference (QPI) measurements on homogeneous terminations of \smb\ revealed the heavy Dirac surface states within the Kondo gap \cite{pirie_imaging_2020, matt_consistency_2020}, establishing \smb\ as the first TKI.

\ptitle{History of YbB$_6$}
\ybb, a second TKI candidate, was predicted to host a mixed valence of Yb around $2.2+$, inverted Yb $4f$ and $5d$ bands, and topological surface states similar to \smb\ \cite{weng_topological_2014}. Indeed, ARPES studies showed an odd number of Fermi pockets on the (001) cleaved surface of \ybb\ \cite{xia_angle-resolved_2014, xu_surface_2014, neupane_non-kondo-like_2015, frantzeskakis_insights_2014}, with the chirality of orbital angular momentum \cite{xia_angle-resolved_2014}, spin texture \cite{xu_surface_2014}, and temperature and photon energy dependence \cite{neupane_non-kondo-like_2015} expected for topological surface states. Quantum oscillation measurements confirmed consistent surface Fermi pocket areas \cite{XiangScience2018, ZhangChinesePhysB2020}. However, magnetic susceptibility \cite{tarascon_magnetic_1980}, optical reflectivity \cite{Nanba_valency_1993}, and x-ray and ultraviolet photoemission \cite{kakizaki_xps_1993} showed that Yb is uniformly divalent in the bulk of \ybb, with a modest Yb$^{3+}$ fraction only at the surface. Furthermore, the highest occupied Yb $4f$ state was consistently found around 1 eV below \Ef\ \cite{kakizaki_xps_1993, xia_angle-resolved_2014, xu_surface_2014, frantzeskakis_insights_2014, neupane_non-kondo-like_2015}, ruling out the TKI scenario, and requiring an alternate explanation for the surface states.

\begin{figure}[htb!] 
\includegraphics{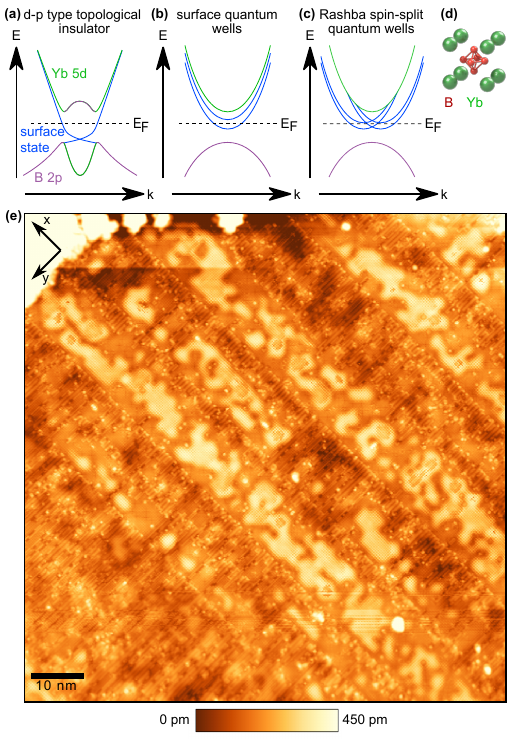}
\caption{(a) Topological insulator, (b) quantum wells, and (c) Rashba spin-split quantum wells band structure proposals. (d) CsCl crystal structure of \ybb. (e) STM topographic image, showing multiple terminations on cleaved \ybb\ (sample bias \Vs = $-0.15$ V; current setpoint \Is = 100 pA). 
Values on the color bar indicate apparent height and not actual height due to the polar surface effects in \ybb.}
\label{fig:cryOver}
\end{figure}

\ptitle{Is YbB$_6$ a topological insulator?} Two distinct mechanisms for the \ybb\ surface states have been proposed.
The first claims strong topological surface states originating from inversion of the bulk Yb $5d$ and B $2p$ bands \cite{neupane_non-kondo-like_2015, chang_two_2015}, as schematized in Fig.\ \ref{fig:cryOver}(a). 
The second claims trivial quantum well states due to band-bending at the polar (001) surface of \ybb\ \cite{kang_electronic_2016, RamankuttyJESRP2016}, as schematized in Fig.\ \ref{fig:cryOver}(b). The latter proposal is supported by the parabolic dispersion of (001) surface states \cite{frantzeskakis_insights_2014}, and additional ARPES measurements on the non-polar (110) surface showing a robust $\sim0.3$ eV gap between the B $2p$ valence and Yb $5d$ conduction bands \cite{kang_electronic_2016}.
Each proposal is supported by density functional theory (DFT) calculations with minor parameter variations. However, neither proposal appears fully consistent with all experimental data. 
All ARPES and quantum oscillation experiments on both (001) and (110) surfaces showed conducting states, despite the various $p$-type, $n$-type, and non-polar terminations expected. Yet no surface channel was seen in magnetoresistance measurements \cite{kim_topological_2014}. Given the likelihood of spectral artifacts from these existing experiments that spatially average over different terminations \cite{matt_consistency_2020}, it remains crucial to conduct local imaging experiments to distinguish between proposals by spatially correlating surface spectral features with specific terminations.

\ptitle{Summary of our accomplishments}
Here, we use STM/STS to demonstrate termination-dependent surface conductivity in \ybb. 
Our spatially-resolved differential conductance spectra (\didv) show strong band bending associated with surface polarity, leading to coexisting insulating and conducting domains. 
The observation of insulating domains on clean regions of \ybb\ definitively rules out the possibility of a strong topological insulator state. 
Within the conducting domains, we observe multiple in-gap peaks reminiscent of van Hove singularities that may stem from Rashba spin-split quantum well states, schematized in Fig.\ \ref{fig:cryOver}(c). 

\begin{figure}[t]  
\includegraphics{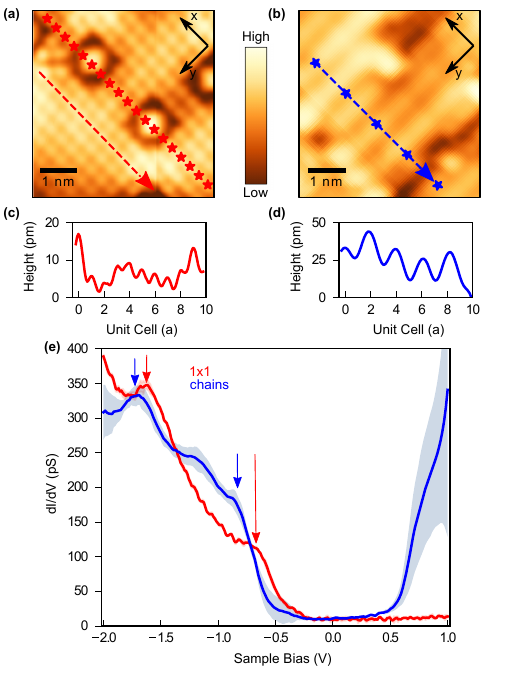}
\caption{(a, b) High resolution topographies of \ybb\ (001) surface showing two distinct terminations:
(a) $1\times1$ domain (\Vs = $-0.4$ V, \Is = 200 pA) and (b) chain domain (\Vs = $0.2$ V, \Is = 100 pA). 
(c, d) Linecuts along respective dashed lines, calibrated to the cryogenic temperature adjusted \ybb\ lattice constant $a=4.1439$ \AA\ \cite{blomberg_single-crystal_1995, EtourneauJSSC1970}. (e) \didv\ spectra averaged over the starred locations in (a) and (b), with shaded standard deviation (\Vs = $-2$ V, \Is = 400 pA, lockin modulation \Vrms = 15 mV).
}
\label{fig:band_bending}
\end{figure}

\ptitle{Crystals and Structural info of YbB$_6$}
Single crystals of YbB$_6$ were grown by the Al-flux method \cite{kim_topological_2014}, cleaved in ultra-high vacuum (UHV) at cryogenic temperatures, and immediately inserted into a home-built STM at 4.2 K.
\ybb, possessing a CsCl-type crystal structure shown in Fig.\ \ref{fig:cryOver}(d), does not have a natural cleavage plane.
Because the intra- and inter-octahedral B-B bonds have similar lengths \cite{blomberg_single-crystal_1995}, several terminations are possible by breaking both types of bonds. 
Breaking inter-octahedral bonds results in the B$_6$ and Yb terminations.
Alternatively, breaking intra-octahedral bonds can lead to disordered and inhomogeneous terminations with incomplete B octahedra.
In our STM study, we find the surface morphology of cleaved \ybb\ exhibits both 1$\times$1 and chain domains, arranged in alternating strips along the [100] crystal axis, as shown in Fig.\ \ref{fig:cryOver}(e). In the 1$\times$1 domains [Fig.\ \ref{fig:band_bending}(a,c)], we measured an atomic spacing consistent with the expected \ybb\ lattice constant of $a=4.1439$ \AA\ \cite{blomberg_single-crystal_1995, EtourneauJSSC1970, lattice_constant}. In the chain domains [Fig.\ \ref{fig:band_bending}(b,d)], the inter-chain period is exactly doubled.

\ptitle{$dI/dV$ spectra, relative band-bending}
Average \didv\ spectra from each domain, which represent the local density of states (DOS), are shown in Fig.\ \ref{fig:band_bending}(e). Both spectra exhibit a large range of reduced DOS around \Ef, and prominent peaks at negative bias, which we identify as the Yb$^{2+}$ $4f$ multiplets. These $4f$ states are often observed by spatially-averaged APRES experiments around $-1$ eV and $-2.3$ eV \cite{xia_angle-resolved_2014, xu_surface_2014, neupane_non-kondo-like_2015, frantzeskakis_insights_2014}, though energies may vary by up to 500 meV between samples \cite{xia_angle-resolved_2014} and time after cleaving \cite{frantzeskakis_insights_2014}.
The chain domain spectra exhibit an additional rise at positive bias, which may be identified as a broad Yb $5d$ band, as observed by inverse photoemission \cite{Nanba_valency_1993}.
The energy shift between our \didv\ spectra suggests terminations with different surface polarities, as commonly observed via spatially-resolved measurements on other semiconductors \cite{tournier-colletta_atomic_2014, matt_consistency_2020}. 
On cleaved \ybb, we expect that a Yb-terminated surface would be positively charged compared to the bulk because the B octahedra take electrons from the Yb layer to form their bonds. 
Bulk free electrons then move towards the surface to screen the positive surface electrostatic potential, forming an accumulation layer with excess electrons ($n$-type), and inducing a downward band-bending. 
Similarly, we expect that a B$_6$-terminated surface would be negatively charged, giving rise to a $p$-type environment and upward band-bending. However, not all B-terminated surfaces need be negatively charged; e.g., the B$_1$ termination requires fewer electrons from the Yb layers to form the octahedral bonds.

\ptitle{$dI/dV$ spectra, relative gap size}
In the \didv\ spectra of Fig.\ \ref{fig:band_bending}(e), the low-conductance energy range around \Ef\ on the chain domain is smaller than that on the 1$\times$1 domain.
The change in energy range can be explained by a combination of tip-induced band bending (TIBB), polar surface band bending, and the presence of in-gap surface states.
STM measurements of insulator gap amplitudes are known to vary from their intrinsic values as a result of poor electronic screening, which allows the electric field generated by the tip to partially penetrate the sample surface \cite{battisti_poor_2017}.
For semiconductors with polar surfaces, tip-induced bending of the bulk bands at the surface can 
lead to apparent variations in gap amplitude and shifts of the gap between different surfaces \cite{tournier-colletta_atomic_2014}. \ybb\ is subject to both band-bending effects, in addition to in-gap states that can alter the electronic screening and thus the overall energy separation.

\begin{figure}[t] 
\includegraphics{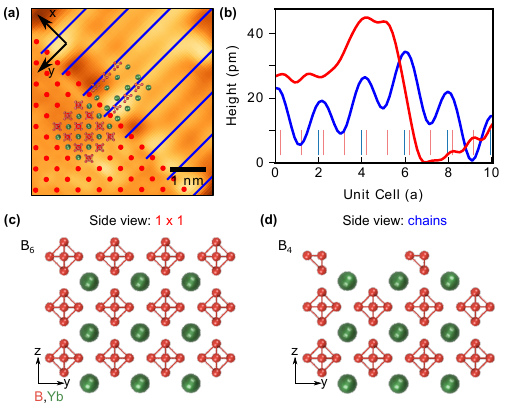}
\caption{(a) Topography showing a boundary between $1\times1$ and chain regions (\Vs = $0.2$ V, \Is = 100 pA).
The blue lines are along the centers of the chains and the red points correspond to the individual atoms on the 1$\times$1 surface.
(b) Averaged linecuts across 1$\times$1 and chain domains in (a). 
Vertical lines represent the unit cell spacing for each surface.
(c, d) Side view of the B$_6$ and B$_4$ surface terminations, respectively.}
\label{fig:interface}
\end{figure}

\ptitle{Inferring surfaces}
Combining topographic and spectroscopic data, we can identify the two observed terminations.
The upward band shift of the 1$\times$1 domain tells us that it must possess a greater number of negatively-charged B atoms with respect to the chain domain, so it cannot represent a Yb termination. Due to its clean structure and atomic periodicity, we infer that the 1$\times$1 domain is either the B$_1$ or B$_6$ termination. In contrast, the double periodicity of the chains tells us that it is a partial layer termination. 
Furthermore, the chains are laterally offset from the 1$\times$1 atoms, as highlighted in Fig.\ \ref{fig:interface}(a), where blue lines indicate the chain centers and red dots indicate the top B atoms of each octahedron. 
Averaged linecuts for the chain and 1$\times$1 domains are shown in Fig.\ \ref{fig:interface}(b). 
We can therefore rule out a partial Yb termination as the chain domain identity, because Yb chains would terminate directly between 1$\times$1 B$_1$ or B$_6$ surface atoms. Therefore, the chains can be identified as a partial B termination.
Due to the up-shifted spectra, we know that the 1$\times$1 domain must have more B atoms than the chain domain, so we conclude that the 1$\times$1 domain must be B$_6$ rather than B$_1$.
The proposed crystal structure for the 1$\times$1 domain and an example B$_4$ termination for the chain domain are illustrated in Fig.\ \ref{fig:interface}(c,d), respectively.
Other possibilities for the chain domain include tilted B$_6$ octahedra, or B$_5$, B$_3$, and B$_2$ terminations.

\begin{figure*}[htb!] 
\includegraphics{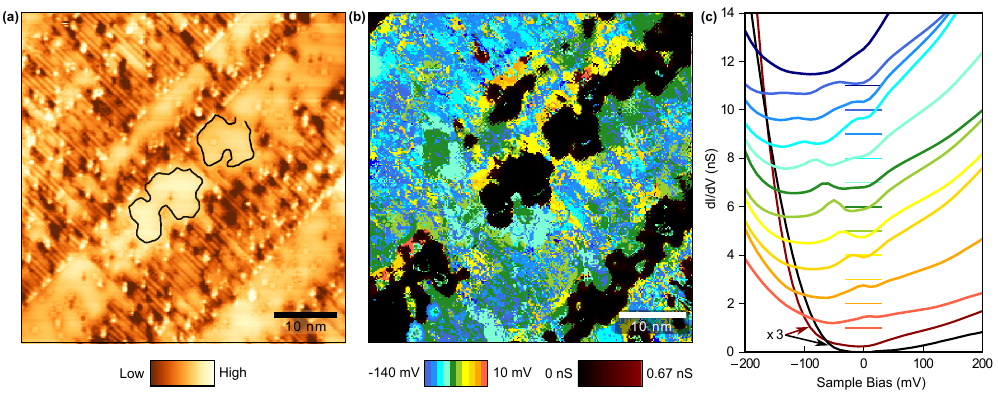}
\caption{ (a) Surface topography, acquired in the same area immediately prior to the \didv\ map that is analyzed in (b,c) (\Vs = $-1$ V, \Is = 50 pA, \Vrms = 15 mV).
(b) Map showing the presence (bright colors) or absence (dark colors) of in-gap states. The energy of the first in-gap peak from the valence band edge is depicted in a rainbow palette. Less than 1\% of the surface exhibits in-gap states without a discernible peak, depicted in dark blue.
Regions without detectable in-gap states are shown instead using a black-red palette depicting the tunneling conductance near \Ef\ (averaged from $-20$ meV to $+20$ meV). 
(c) Averaged \didv\ spectra from (b) showing sub-gap peaks in the 10 color bins, each offset vertically by 1 nS with horizontal lines to represent their zero for clearer comparison.
Close to 25\% of the surface spectra show no in-gap states. These gapped spectra, averaged over regions with tunneling conductance below 1 pS (black) and above 1 pS (dark red) are shown with zero offset, but $3\times$ magnification to emphasize the true Fermi-level gap.
}
\label{fig:rashba}
\end{figure*}

\ptitle{YbB$_6$ is not a TI}
To understand the \ybb\ surface states previously observed by spatially-averaging probes \cite{xia_angle-resolved_2014, xu_surface_2014, neupane_non-kondo-like_2015, frantzeskakis_insights_2014, XiangScience2018, ZhangChinesePhysB2020}, we zoom in on the near-\Ef\ states that are swamped by the dominant $4f$ multiplets in Fig.\ \ref{fig:band_bending}(e). 
Figure \ref{fig:rashba} shows a topography and \didv\ map of the same area over a smaller energy range around \Ef, acquired with lower junction resistance for increased sensitivity to sub-gap features. 
With this increased sensitivity, we observe a true insulating gap on $\sim25\%$ of the surface, corresponding to clean 1$\times$1 regions in Fig.\ \ref{fig:rashba}(a), and depicted in a black-red colorscale in Fig.\ \ref{fig:rashba}(b). 
The average spectra from these regions are shown in corresponding colors in Fig.\ \ref{fig:rashba}(c), with apparent gap amplitude $\sim100$ meV roughly consistent with the known bulk gap \cite{tarascon_magnetic_1980}. We caution that TIBB is likely even more significant at this lower junction resistance where the tip is closer to the sample, but we emphasize that
TIBB cannot create an apparent gap, though it can change an existing gap amplitude. Therefore, the ${\sim}25\%$ gapped area definitively rules out a strong topological surface state, which would necessarily cover all surfaces \cite{fu_topological_2007b, fu_topological_2007}.

\ptitle{Quantum well states}
The remaining ${\sim}75\%$ of the \ybb\ surface, comprising most of the chain domains and a smaller fraction of the 1$\times$1 domains, shows in-gap states with one or two peaks at energies near \Ef. 
The average spectra, binned by their minimum in-gap peak energy, are shown in Fig.\ \ref{fig:rashba}(c). 
Though \ybb\ is a bulk insulator \cite{tarascon_magnetic_1980}, the demonstrated surface polarity may bend a bulk band to cross \Ef\ at the surface. Confinement within a few nanometers of the surface reduces the 3-dimensional band to a discrete number of 2-dimensional bands known as quantum well states (QWS), as suggested by previous ARPES experiments \cite{frantzeskakis_insights_2014, kang_electronic_2016}, and schematized in Fig.\ \ref{fig:cryOver}(b). But these QWS alone would not give rise to a peak in the density of states. To explain our observed peak, as well as the spin-polarization measured by ARPES \cite{xu_surface_2014}, we suggest that Rashba spin-orbit coupling (SOC) splits the surface bands, as schematized in Fig.\ \ref{fig:cryOver}(c). The spin-splitting would give rise to a van Hove singularity (vHS) and observable peak in the density of states at the minimum energy of the QWS \cite{tournier-colletta_atomic_2014, ast_local_2007}. 
The ${\sim}150$ mV variation of sub-gap peak energy in Fig.\ \ref{fig:rashba}(c) may be attributed to band bending from the two surface terminations, and further local band bending from the many atomic-scale defects apparent in Fig.\ \ref{fig:rashba}(a). 

\ptitle{Conclusion} 
In conclusion, our STM/STS study provides the first real-space imaging of \ybb.
On the (001) surface, we observe two different atomic terminations whose distinct polarity is evidenced by spectra shifts greater than 100 meV. Furthermore, we observe both conducting and insulating nanoscale domains, definitively ruling out the proposed strong topological surface state for \ybb.
Instead, we show that the surface states previously observed by spatially-averaging techniques arise from quantum well states induced by surface band bending. 
We further show that the quantum well states are Rashba spin-split, evidenced by sub-gap peaks consistent with van Hove singularities in the conductance spectra. 
The coexisting nanoscale domains with differing surface polarities suggests \ybb\ may represent a new class of strongly-correlated materials with utility for spin-polarized $p$-$n$ junctions or other spintronic devices \cite{tournier-colletta_atomic_2014}.

\begin{acknowledgments}
We acknowledge Ruizhe Kang, Harry Pirie, Christian Matt, Can-Li Song, and Dennis Huang for helpful discussions. A.C. was funded by the Air Force Office of Scientific Research Multidisciplinary University Research Initiative through Award No. FA9550-21-1-0429.  
\end{acknowledgments}

%

\end{document}